\documentclass[conference]{IEEEtran}
\IEEEoverridecommandlockouts

\usepackage{cite}
\usepackage{amsmath,amssymb,amsfonts}
\usepackage{algorithmic}
\usepackage{graphicx}
\usepackage{textcomp}
\usepackage{xcolor}
\usepackage[left=0.673in,right=0.673in,top=0.7in,bottom=0.967in]{geometry}
\def\BibTeX{{\rm B\kern-.05em{\sc i\kern-.025em b}\kern-.08em
    T\kern-.1667em\lower.7ex\hbox{E}\kern-.125emX}}    
\begin{document}

\title{Auction-Based RIS Allocation With DRL: Controlling the Cost-Performance Trade-Off
\thanks{This research was funded in whole or in part by the
Austrian Science Fund (FWF) 10.55776/PAT4490824. For
open access purposes, the author has applied a CC BY public
copyright license to any author accepted manuscript version
arising from this submission.}
}

\author{\IEEEauthorblockN{1\textsuperscript{st} Martin Mark Zan}
\IEEEauthorblockA{\textit{Institute of Telecommunications} \\
\textit{TU Wien}\\
Vienna, Austria \\
martin.zan@tuwien.ac.at}
\and
\IEEEauthorblockN{2\textsuperscript{nd} Stefan Schwarz}
\IEEEauthorblockA{\textit{Institute of Telecommunications} \\
\textit{TU Wien}\\
Vienna, Austria \\
stefan.schwarz@tuwien.ac.at}
}

\maketitle
\begin{minipage}{500pt}
\vspace{-400pt}
\centering{
\footnotesize{© 2026 IEEE. Personal use of this material is permitted. This is the author's version of the work accepted for publication in IEEE International Conference on Communications (ICC 2026).}}
\end{minipage}

\begin{abstract}
We study the allocation of reconfigurable intelligent surfaces (RISs) in a multi-cell wireless network, where base stations compete for control of shared RIS units deployed at the cell edges. These RISs, provided by an independent operator, are dynamically leased to the highest bidder using a simultaneously ascending auction format. Each base station estimates the utility of acquiring additional RISs based on macroscopic channel parameters, enabling a scalable and low-overhead allocation mechanism. To optimize the bidding behavior, we integrate deep reinforcement learning (DRL) agents that learn to maximize performance while adhering to budget constraints. Through simulations in clustered cell-edge environments, we demonstrate that reinforcement learning (RL)-based bidding significantly outperforms heuristic strategies, achieving optimal trade-offs between cost and spectral efficiency. Furthermore, we introduce a tunable parameter that governs the bidding aggressiveness of RL agents, enabling a flexible control of the trade-off between network performance and expenditure. Our results highlight the potential of combining auction-based allocation with adaptive RL mechanisms for efficient and fair utilization of RISs in next-generation wireless networks.
\end{abstract}


\section{Introduction}

\IEEEPARstart{I}{n} the pursuit of beyond-5G and 6G wireless networks, improving spectral and energy efficiency has become a central research objective. Reconfigurable intelligent surfaces (RIS) have recently emerged as a promising technology to reshape the radio environment and enhance coverage in challenging propagation conditions \cite{Zeng2021}. By enabling programmable reflection of incident signals, RISs provide a cost- and energy-efficient means to strengthen desired links and suppress interference, thereby complementing conventional base station and user equipment capabilities \cite{Msleh2023, Le2021}. Their practical use, however, raises fundamental design questions, including how to deploy RISs in realistic network topologies and how to efficiently allocate them to different transmitters and users.

In this work, we concentrate on scenarios where multiple base stations compete for control of RISs located near the cell boundary. Such situations naturally arise at the cell edge, where the positioning of RISs can have a decisive impact on system performance. This setting introduces a resource allocation problem: when several transmitters contend for the same RISs, an effective mechanism is required to determine access in a fair and efficient manner. To address this challenge, we consider a market-inspired approach in which RISs are not permanently assigned but are dynamically leased to the highest bidder through an ascending auction format. This mechanism provides a low-complexity and scalable alternative to combinatorial allocation methods \cite{Nisan2001}, while capturing the competitive interactions between base stations. Similar auction-based approaches have been successfully applied in spectrum allocation, where simultaneous ascending auctions enable efficient and competitive distribution of scarce frequency resources \cite{Cramton2017, Milgrom2004}. A related approach was recently investigated in \cite{Schwarz2024}, where an auction-based RIS allocation was studied in a multi-operator setting. Building on this framework, we further employ reinforcement learning (RL) to optimize the bidding strategies of individual agents. By doing so, the agents can adapt their behavior to the evolving auction environment, selectively targeting high-value RISs and avoiding inefficient bidding. Our study demonstrates that this combination of auction-based allocation and RL-driven bidding achieves superior trade-offs between cost and performance compared to heuristic approaches.

\textit{Notation:} The complex Gaussian distribution with mean $\mu$ and covariance matrix $C$ is denoted as $\mathcal{CN}(\mu,C)$. The uniform distribution over the interval $[a,b]$ is written as $\mathcal{U}(a,b)$. The transpose and Hermitian transpose of a vector $\mathbf{x}$ are $\mathbf{x}^T$ and $\mathbf{x}^H$, respectively, and its $i$-th element is $\mathbf{x}[i]$. The Euclidean norm of $\mathbf{x}$ is $\|\mathbf{x}\|$. The size of a set $\mathcal{X}$ is $|\mathcal{X}|$, and the empty set is $\emptyset$. The expectation of a random variable $r$ is $\mathbb{E}[r]$, and the phase of a complex number $z$ is $\arg(z)$.

\section{System model}\label{II}

Let us have $N_\text{BS}$ base stations which serve $N_\text{UE}$ users in parallel over $N_\text{RIS}$ RISs. We assume that each base station is equipped with $M_\text{BS}$ antennas and users are equipped with a single antenna, while each RIS consists of $M_\text{RIS}$ reconfigurable elements.

We have a direct channel between each base station $b$ and each user $u$, denoted as $\mathbf{h}_{u,b}^{\text{direct}} \in \mathbb{C}^{M_{\text{BS}} \times 1}$. This channel is written as
\begin{equation}
    \mathbf{h}_{u,b}^{\text{direct}} = \gamma_{u,b} \mathbf{g}_{u,b}.
\end{equation}
The channel gain $\gamma_{u,b}$ depends on the distance between the user and the base station. The fading component $\mathbf{g}_{u,b}$ follows a zero-mean unit-variance complex Gaussian distribution, and $\mathbb{E}\bigl[\left\|\mathbf{g}_{u,b}\right\|^2\bigr] = M_\text{BS}$. The normalization condition on the channel is $\mathbb{E} \bigl[ \| \mathbf{h}^\text{direct}_{u,b} \|^2 \bigr] = \gamma_{u,b}^2 M_\text{BS}$. We assume that the the base station to user link is non-line-of-sight (NLOS) and it is strongly shadowed, so that the line-of-sight (LOS) component can be neglected. This makes sense in this setup, since we are interested in RIS-aided transmissions.

In addition to this direct channel, we have RIS-assisted channels. These are composed of the channel between each base station and the RISs, the RISs' responses, and the channel between the RISs and the users. 

Let us consider the channel between the base station $b$ and RIS $r$: $\mathbf{H}_{r,b} \in \mathbb{C}^{M_\text{RIS} \times M_\text{BS}}$. This is a matrix valued channel, containing all channels between $M_\text{BS}$ base station antennas and the $M_\text{RIS}$ RIS elements. It is written as:
\begin{equation}
    \mathbf{H}_{r,b} = \gamma_{r,b}\, \mathbf{a}(\psi_{r,b}) \mathbf{a}(\theta_{r,b})^T.
\end{equation}
Here, $\gamma_{r,b}$ is the path gain between the base station and the RIS, $\mathbf{a}(\psi_{r,b})$ denotes the directional RIS response vector, and $\mathbf{a}(\theta_{r,b})$ is the directional base station response vector \cite{Balanis2005}. They depend on the angle-of-arrival $\psi_{r,b}$ of the signal at the RIS and the angle-of-departure $\theta_{r,b}$ of the signal at the base station. For simplicity, it is assumed that the base station to RIS channel is under strong LOS, so that the fading NLOS component is neglected. This assumption is reasonable, since RISs are typically deployed in locations that ensure unobstructed visibility to nearby base stations, for example, mounted on building facades or rooftops in urban environments.

The channel between RIS $r$ and user $u$ is $\mathbf{h}_{u,r} \in \mathbb{C}^{M_{\text{RIS}} \times 1}$. We use a Rician channel model, in which the channel is composed of a non-fading LOS component and a random Gaussian fading NLOS component. The relative strength of the two channel components is characterized by the Rician $K$-factor, which is higher under LOS conditions and lower under NLOS conditions. It is written as follows
\begin{equation}
    \mathbf{h}_{u,r} = \gamma_{u,r} \left(\sqrt{\frac{K_{u,r}}{1+K_{u,r}}} \mathbf{a}(\theta_{u,r}) + \sqrt{\frac{1}{1+K_{u,r}}} \mathbf{g}_{u,r}  \right).
\end{equation}
The channel gain $\gamma_{u,r}$ depends on the line-of-sight conditions of the link (higher gain under line-of-sight, lower under non-line-of-sight) and on the distance between the user and the RIS. The non-fading component $\mathbf{a}(\theta_{u,r})$ is the directional base station response vector \cite{Balanis2005}, and $\mathbb{E}\bigl[\left\|\mathbf{a}(\theta_{u,r})\right\|^2\bigr] = M_\text{RIS}$. The angle-of-departure of the signal at the RIS is $\theta_{u,r}$. Let us define $ k_{u,r} = \sqrt{\frac{K_{u,r}}{1+K_{u,r}}}$ and $ \bar{k}_{u,r} = \sqrt{\frac{1}{1+K_{u,r}}}$. The vector $\mathbf{g}_{u,r}$ is defined analogously as in the direct channel.

The RIS infrastructure is managed by an independent provider who leases the units on-demand to the highest bidder. Once a base station gains control of a RIS, it can configure its phase-response to enhance signal quality for its associated users. We consider a diagonal RIS, so that the RIS response can be characterized by a diagonal phase-shift matrix $\boldsymbol{\Phi}_r = \mathrm{diag}(e^{j\phi_{r,1}},\ldots,e^{j\phi_{r,M_\text{RIS}}}) \in \mathbb{C}^{M_{\text{RIS}} \times M_{\text{RIS}}}$. With this, the overall RIS-assisted channel with all RISs $r$ is $\mathbf{h}_{u, b}^\text{indirect} \in \mathbb{C}^{M_{\text{BS}} \times 1}$, which is defined as:
\begin{equation}
    \mathbf{h}_{u, b}^\text{indirect} = \sum_{r=1}^{N_\text{RIS}} (\mathbf{h}_{u,r}^T \boldsymbol{\Phi}_r \mathbf{H}_{r,b})^T.
\end{equation}

For RIS $r$ with elements indexed by $i$, the optimized phase-shift is given by
\begin{equation}
\phi_{r,i} = - \left( \arg\big(\mathbf{a}(\theta_{u,r})[i]\big) + \arg\big(\mathbf{a}(\theta_{r,d})[i]\big) \right).
\end{equation}
We assume that random scattering components $\mathbf{g}_{u,r}$ change so quickly that we cannot estimate them reliably. Therefore, they cannot be phase-aligned, and they contribute incoherently. If a RIS is not assigned to the serving base station, its elements are modeled as random phases, i.e., $\phi_{r,i} \sim \mathcal{U}(0, 2\pi)$.

Furthermore, the total direct + assisted channel $\mathbf{h}_{u,b} \in \mathbb{C}^{M_\text{BS} \times 1}$ is
\begin{equation}
    \mathbf{h}_{u,b} = \mathbf{h}_{u,b}^{\text{direct}} + \mathbf{h}_{u,b}^{\text{indirect}}.
\end{equation}

We consider a system model in which each base station serves multiple users using time-orthogonal resources, such that only one user is scheduled per time slot. Based on these assumptions, the frequency-flat single-input single-output (SISO) downlink input-output relation for user $u$ can be expressed as follows:
\begin{equation}
    y_u = \mathbf{h}_{u,d}^T \mathbf{f}_{u,d} x_{u} +\sum\nolimits_{b \neq d} \mathbf{h}_{u,b}^T \mathbf{f}_{j_b,b} x_{j_b} + n_u, 
\end{equation}
where $d$ denotes the index of the base station serving user $u$, $\mathbf{f}_{u,d} \in \mathbb{C}^{M_\text{BS} \times 1}$ is the beamforming vector applied to transmit to user $u$ at base station $d$. $P_{u,b}$ is the power used to transmit to user $u$ from base station $b$ and is considered as part of the beamformer, such that $\mathbb{E} \bigl[ \left\|\mathbf{f}_{u,b}\right\|^{2}\bigr] = P_{u,b}$. $x_u$ is the symbol intended for user $u$; normalization is applied to transmit symbols, i.e. $\mathbb{E}\bigl[|x_u|^2\bigr] = 1$. $j_b$ denotes the index of the user served by base station $b$ and $n_u \sim \mathcal{CN}(0, \sigma^2_n)$ is noise. We also assume that the beamforming is based only on the directional parts of the channel, the angle-dependent array steering vectors, since the NLOS part changes too quickly to be reliably estimated.

The users' instantaneous signal to interference and noise ratios (SINRs) $\beta^{(d)}_u$ and achievable rates $r^{(d)}_u$ are:
\begin{equation}
    \beta^{(d)}_u = \frac{|\mathbf{h}_{u,d}^T \mathbf{f}_{u,d}|^{2}}{\sigma^2_n + \sum\nolimits_{\substack{b \neq d}}{|\mathbf{h}_{u,b}^T \mathbf{f}_{j_b,b}|}^{2}},\quad r^{(d)}_u = \text{log}_2(1+\beta^{(d)}_u). \label{eq:instant}
\end{equation}

\section{Utility Estimation}

In order to allocate RISs, each base station needs to evaluate the potential performance gain that can be achieved when certain RISs are assigned to it. Since perfect channel state information is not available before the RISs are configured, we rely on macroscopic channel properties to obtain coarse estimates of the achievable SINR and sum rates. These estimates are then used to define a utility function which guides the allocation.

\subsection{SINR Estimation}

Based on the channel model in Section~\ref{II}, we estimate the users' SINR using the macroscopic channel parameters, by approximating the instantaneous power values in \eqref{eq:instant} with their expected values. For sufficiently large antenna arrays and RISs, this approximation provides a good estimate due to the law of large numbers. We decompose the received powers into direct, coherent RIS-assisted, and incoherent RIS-assisted components, as well as direct and RIS-assisted interference. The estimated SINR for user $u$ served by base station $d$ is

\begin{equation}
\hat{\beta}^{(d)}_u = \frac{p_d + p_c + p_i}{\sigma_n^2 + i_d + i_i},
\end{equation}
where
\begin{align*}
p_d &= \gamma_{u,d}^2 P_{u,d}, \\
p_c &= \left(\sum_{r \in R^{(d)}} \gamma_{u,r} \gamma_{r,d} k_{u,r} 
       \sqrt{\tfrac{P_{u,d} M_{\text{BS}}}{|R^{(d)}|}} \, M_{\text{RIS}} \right)^2, \\
p_i &= \sum_{r \in R^{(d)}} \gamma_{u,r}^2 \gamma_{r,d}^2 \bar{k}_{u,r}^2 
       \tfrac{P_{u,d} M_{\text{BS}}}{|R^{(d)}|} M_{\text{RIS}}, \\
i_d &= \sum_{b \neq d} \gamma_{u,b}^2 P_{j_b,b}, \\
i_i &= \sum_{b \neq d} \sum_{r \notin R^{(d)}} \gamma_{u,r}^2 \gamma_{b,r}^2 
       P_{j_b,b} M_{\text{RIS}}.
\end{align*}
Here $R^{(d)}$ denotes the set of RISs assigned to base station $d$. The numerator consists of the direct signal $p_d$, the coherent RIS-assisted components $p_c$, and the incoherent RIS-assisted components $p_i$. The denominator includes noise $\sigma_n^2$, direct interference $i_d$ from non-serving base stations $b \neq d$, and RIS-assisted interference $i_i$ caused by RISs not assigned to the serving base station $d$.

We assume that the beamformer of each base station points towards its assigned RISs with uniform power distribution. If no RIS is assigned, the beamformer is chosen to be isotropic. In this case, we only have a channel with $K$-factor equal to 0. Since this random Rayleigh fading channel is not known to the base station, we cannot match the beamformer to this channel and therefore can only select it randomly. Interfering base stations are also assumed to employ isotropic beamforming, since their RIS assignments are unknown to base station $d$. Finally, we assume that the directional base station response vectors are asymptotically orthogonal. The estimated achievable rate then follows as
\begin{equation}
\hat{r}^{(d)}_u = \log_2 \left(1 + \hat{\beta}^{(d)}_u \right).
\end{equation}

\subsection{Utility Function}

We adopt a utility function that measures the percentage sum-rate improvement compared to the case without RISs. The utility for a given allocation $R^{(b)}$ is defined as

\begin{equation}
U^{(b)}(R^{(b)}) = \frac{\sum_{u=1}^{N^{(b)}_{\text{UE}}} \hat{r}^{(b)}_u(R^{(b)})}{\sum_{u=1}^{N^{(b)}_{\text{UE}}} \hat{r}^{(b)}_u(\emptyset)} - 1,
\end{equation}
where $\hat{r}^{(b)}_u(R^{(b)})$ denotes the estimated rate of user $u$ when RISs $R^{(b)}$ are allocated \cite{Schwarz2024}.

\subsection{Auction Format}
The allocation of RISs to base stations is performed using a simultaneously ascending (``Japanese'' forward) auction format, as introduced in \cite{Schwarz2024}. This mechanism was originally motivated by the need for a low-complexity alternative to combinatorial auctions such as the Vickrey-Clarke-Groves (VCG) mechanism \cite{Nisan2001}. In each round $t$, the auctioneer increases the price $p_t$ of each RIS by a fixed increment $\Delta_p = p_t - p_{t-1}$. Base stations then indicate their willingness to pay the current price for individual RIS units using a binary "bid" vector $b_t^{(b)} \in \{0,1\}^{N_\text{RIS}}$. If a RIS receives only one bid, it is allocated to the corresponding base station; if a RIS receives more than one bid, the bidding continues in the next round; if the RIS receives zero bids then it remains unassigned, and it's phase-response is set randomly. To ensure consistency, an activity rule is enforced, i.e., a base station cannot rejoin the bidding for a RIS if it did not bid for it in the previous round \cite{Roughgarden2016}. We adopt this mechanism in our setup without modification.

\section{Bidding Strategies}

At each auction round $t$, base stations must decide which RISs are worth bidding for, based on the current price $p_t$ and their estimated utility. Let $R_{t-1}^{(b)}$ denote the set of RISs already allocated in previous rounds to base station $b$. The remaining unassigned RISs form the set $R_t = \{1, \ldots, N_\text{RIS}\} \setminus \bigcup_b R_{t-1}^{(b)}$. Each base station $b$ then evaluates the potential gain of adding an individual RIS $r \in R_t$ to its current allocation $R_{t-1}^{(b)}$.

Ideally, this decision would be based on the expected utility of all possible combinations of the remaining RISs. However, such a combinatorial evaluation becomes infeasible when the number of RISs is large, due to the exponential number of subsets. To keep the computation tractable, we adopt a simplified approach, where each base station evaluates the benefit of acquiring a single additional RIS $r$, assuming no other RIS will be acquired in the same round. The estimated marginal value of RIS $r$ is then given by:

\begin{equation}
V_t^{(b)}(r) = U^{(b)}(R_{t-1}^{(b)} \cup \{r\}) - U^{(b)}(R_{t-1}^{(b)}).
\end{equation}

To standardize the scale of utility gains across different bidding scenarios, the estimated values are normalized as:

\begin{equation}
V_t^{(b)}(r) \leftarrow \frac{V_t^{(b)}(r)}{\max_{r'} \left| V_t^{(b)}(r') \right| },\label{eq:value}
\end{equation}
resulting in values bounded within the interval $[-1, 1]$. A positive value indicates a potential improvement in performance, while a negative value suggests a possible degradation. Negative values can arise due to our assumption of uniform power allocation over RIS-directed transmit beams at the base station.

Based on these estimated values, different bidding strategies can be applied. In the following subsections, we consider three approaches: two heuristic strategies based on dominant bidding behavior, and a reinforcement learning based strategy that adapts over time to maximize long-term reward.

\subsection{Heuristic Bidding}\label{IV-A}

As a baseline approach, we implement a simple greedy bidding strategy. At each auction round $t$, base station $b$ computes the value $V_t^{(b)}(r)$ for each available RIS $r \in R_t$, based on the definition in \eqref{eq:value}. After normalization, the RISs are ranked in descending order of estimated value.

To determine how many RISs to bid for, we calculate the remaining budget $B^{(b)}_t$ as the difference between the initial budget $B^{(b)}_0$ and the total cost already paid for previously assigned RISs $R^{(b)}_{t-1}$. A conservative bidding rule is then applied: the base station places bids for the $\left\lfloor B^{(b)}_t / p_t \right\rfloor$ most valuable RISs at round $t$. Although this strategy does not account for long-term optimization or strategic interaction across rounds, it provides a low-complexity and deterministic baseline that reflects greedy decision-making based solely on immediate value.

In addition to this value-based approach, we also evaluate a simpler distance-based heuristic. In this variant, the value function is replaced with a distance-derived metric defined as
\begin{equation}
VD_t^{(b)}(r) = \frac{1}{\text{dist}(b, r) + \varepsilon},
\end{equation}
where $\text{dist}(b, r)$ is the Euclidean distance between base station $b$ and RIS $r$, and $\varepsilon$ is a small constant to avoid invalid division. The rest of the bidding procedure remains unchanged. This strategy assumes that proximity correlates with utility, and serves as a lower-complexity baseline that does not rely on SINR estimation.

\subsection{RL-Based Bidding}

Alternatively to these heuristic strategies, we explore the use of RL to derive optimized bidding strategies. Each base station operates an independent RL agent, trained to make decisions based only on its local observations, without coordination or information sharing among other base stations. In the following, we define the structure of the environment used for training these agents, including the state and observation spaces, action representation, and the reward function.

\subsubsection*{a) States}

The complete state of the auction environment at time step $t$ is defined as:
\begin{equation}
S_t = \left( p_t, \left\{ V^{(b)}_t(r), B^{(b)}_t \ \middle| \ \forall b, r \right\} \right).
\end{equation}

To ensure a fixed-length state representation compatible with RL algorithms, we assign $V^{(b)}_t(r) = 0$ for all RISs $r$ that:
\begin{itemize}
    \item are no longer available (i.e., already assigned),
    \item were not bid on by base station $b$ in the previous round (in line with the enforced activity rule),
    \item or for which the current price $p_t$ exceeds the remaining budget of base station $b$.
\end{itemize}

This approach avoids explicitly tracking previously assigned RISs, while preserving the same semantics. It also enables scalability to environments with different numbers of RISs, as unused RIS slots can be zero-padded. The environment maintains this full state internally, although each agent operates only on its local observation, reflecting the partially observable nature of the problem.

\subsubsection*{b) Observations}

The observation available to base station $b$ at time step $t$ is given by:
\begin{equation}
O^{(b)}_t = \left( p_t, B^{(b)}_t, \left\{ V^{(b)}_t(r) \ \middle| \ \forall r \right\} \right).
\end{equation}

\subsubsection*{c) Actions}

At each time step $t$, based on its observation $O^{(b)}_t$, each agent (i.e., base station $b$) outputs a bid vector $b^{(b)}_t \in \{0, 1\}^{N_{\text{RIS}}}$,
where the binary entry $b^{(b)}_t[r] = 1$ indicates that the base station is willing to bid for RIS $r$ at the current price $p_t$, and $b^{(b)}_t[r] = 0$ otherwise.

\subsubsection*{d) Rewards}

The reward function is designed to guide each agent toward maximizing the value of its bids while staying within budget and avoiding wasteful spending. It consists of three components:

\begin{itemize}
    \item R1 (value of bids): This term measures the total value of the RISs the agent bids on in a given round. It is computed as the sum of the estimated values $V^{(b)}_t(r)$ for all RISs $r$ for which $b^{(b)}_t[r] = 1$:
    \begin{equation}
    R1 = \sum_{r = 1}^{N_{\text{RIS}}} V^{(b)}_t(r) \cdot b^{(b)}_t[r].
    \end{equation}
    
    \item R2 (cost of bids): This term penalizes agents for the total cost of their bids, scaled by the bid intensity parameter $\beta$. The cost is computed as the number of bids placed multiplied by the current price:
    \begin{equation}
    R2 = \beta \cdot p_t \cdot \sum_{r = 1}^{N_{\text{RIS}}} b^{(b)}_t[r].
    \end{equation}
    
    \item R3 (overspending penalty): This component introduces an additional penalty if the total bid cost exceeds the agent’s remaining budget. The penalty is proportional to the overshoot and scaled by $2 \beta$:
    \begin{equation}
    R3 =2 \beta \cdot \max\left( p_t \cdot \sum_{r = 1}^{N_{\text{RIS}}} b^{(b)}_t[r] - B^{(b)}_t,\ 0 \right).
    \end{equation}
    The scaling factor of 2 in R3 is chosen to make overspending penalties clearly distinguishable from the regular cost term R2. This ensures that the agent perceives budget violations as more severe than ordinary expenditures, while avoiding excessively high penalties that would lead to overly conservative bidding behavior.
\end{itemize}

The final reward is then computed as:
\begin{equation}
r^{(b)}_t = R1 - R2 - R3
\end{equation}

This reward formulation encourages the agent to bid on high-value RISs (via R1), penalizes excessive or unaffordable bidding (via R2), and strictly discourages budget violations (via R3). Notably, the reward is computed before the actual auction outcome, allowing the agent to receive immediate feedback on its decisions and learn more efficiently. This leads to a denser and more informative reward signal compared to outcome-based rewards. To enable generalization across environments with different budgets and prices, both the price $p_t$ and the budget $B^{(b)}_t$ are normalized by the initial budget. The bid intensity parameter $\beta$ is selected from a range of representative values to illustrate the cost-performance trade-off, and the values used in Section~\ref{V} are found to yield stable learning behavior and meaningful operating points.

We also evaluated several alternative reward structures. First, we tested using only the actual performance improvement (i.e., post-auction value gain), along with the cost and overspending penalty computed from the RISs that were actually won. However, the delayed feedback significantly hindered learning: agents were unable to reliably identify high-value RISs. Second, we experimented with nonlinear penalty functions in R3. Both quadratic and exponential penalties led to numerical instability: large bid costs produced disproportionately large negative rewards, causing the learning process to collapse.

\subsubsection*{e) Implementation}
We implemented our RL-based auction environment using Gymnasium 1.1.1 \cite{gymnasium} and PettingZoo 1.25.0 \cite{pettingzoo} for multi-agent support, along with SuperSuit 3.10.0 \cite{supersuit} for preprocessing and environment wrapping. Agents were trained using the Stable-Baselines3 2.6.0 \cite{sb3} implementation of the Proximal Policy Optimization (PPO) algorithm \cite{ppo}, with default hyperparameters.

\begin{figure}[t]
\centerline{\includegraphics[width=\columnwidth]{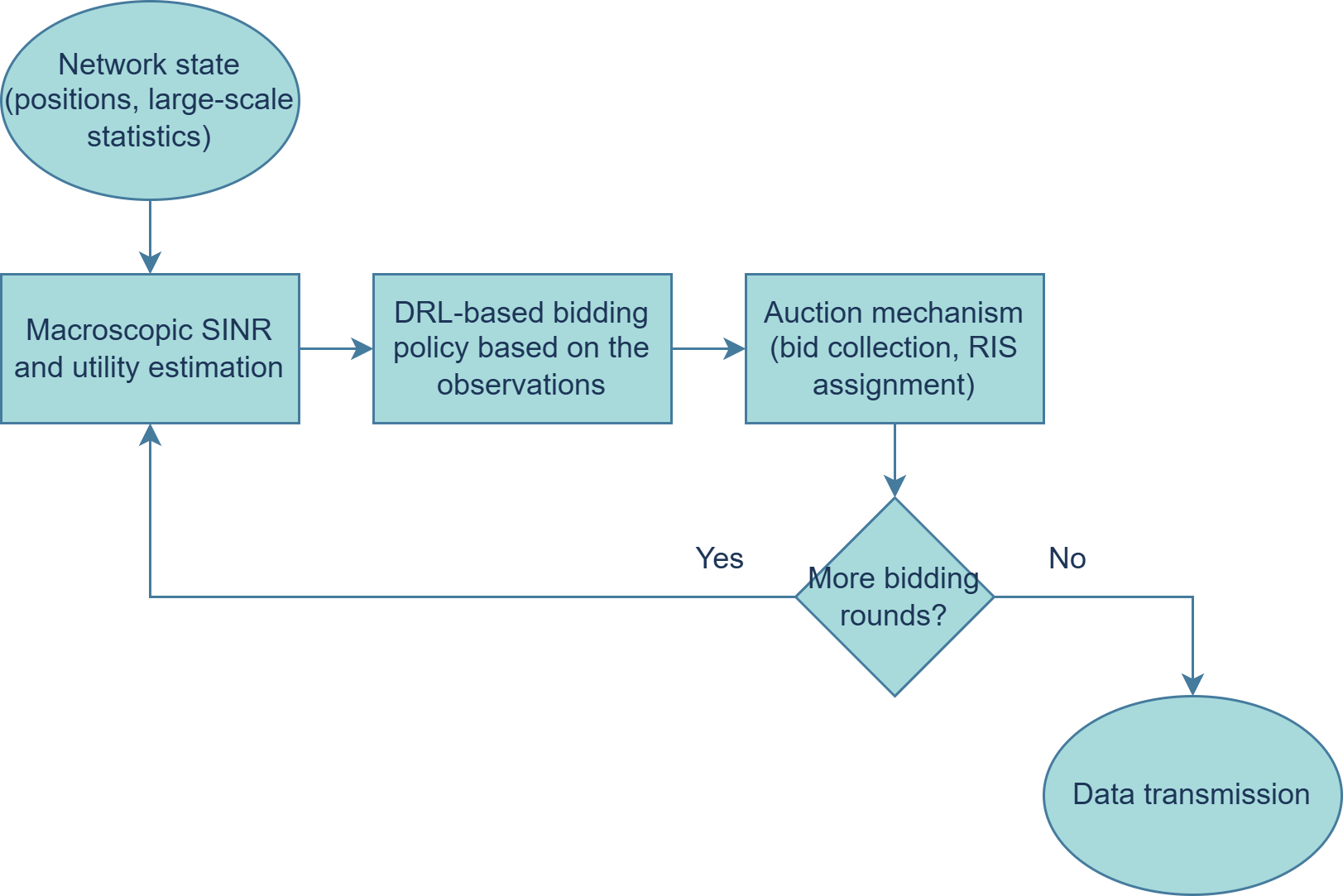}}
\caption{System workflow of the proposed auction-based RIS allocation framework. 
Macroscopic SINR and utility estimates are used as inputs to a DRL-based bidding policy, 
which generates bids for the auction mechanism. The auction iterates over bidding rounds 
until a termination condition is met, after which data transmission 
is performed.}
\label{fig:workflow}
\end{figure}

The environment supports a variable number of users, base stations, and RISs. Training is conducted episodically, where each episode corresponds to a full auction from start to finish. The episode length is variable and depends on the number of RISs and the bidding behavior of the agents. After each episode, a new environment is generated with randomized channel realizations and geometry. In practical scenarios with time-varying user locations, the trained bidding policy can be re-executed to quickly reassign RIS resources, without requiring retraining of the model. The system workflow can be seen on Fig.~\ref{fig:workflow}.

The implementation and the full list of dependencies and versions will be made available upon acceptance of the work at: https://github.com/MartinMarkZan.

\section{Simulations} \label{V}

\begin{figure}[t]
\centerline{\includegraphics[width=\columnwidth]{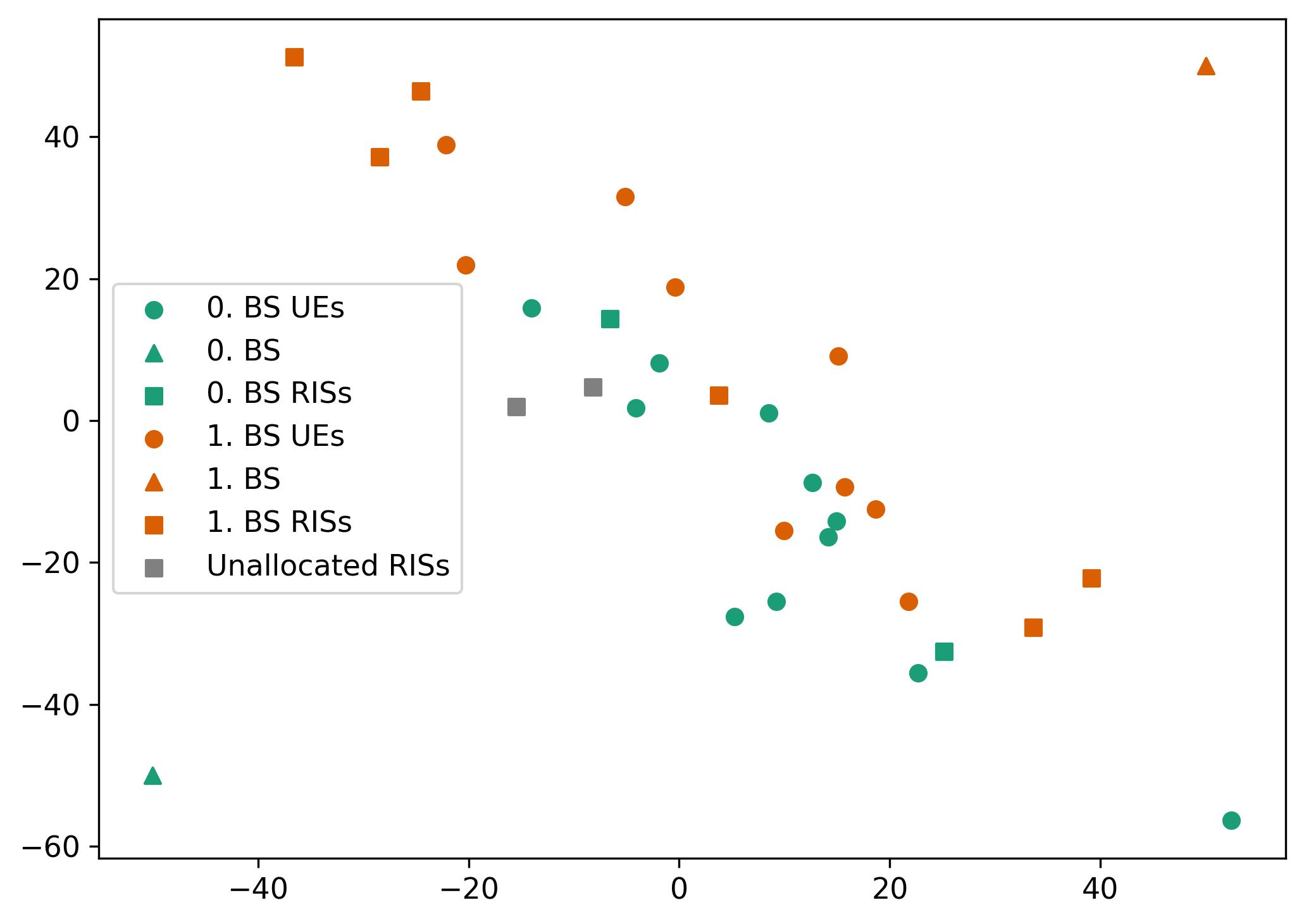}}
\caption{Simulation geometry with two base stations (BS) at opposite ends, clustered users (UE), and RISs positioned along the cell boundary. The figure shows an example allocation of RISs obtained using a reinforcement learning agent with bid intensity parameter $\beta=3$.}
\label{fig:plot_positions_RL}
\end{figure}

\begin{table}[t]
\caption{Simulation Parameters}
\begin{center}
\begin{tabular}{|c|c|}
\hline
Carrier frequency & $f_c = 26\,\text{GHz}$ \\
\hline
Number of base stations & $N_\text{BS} = 2$ \\
\hline
Number of base station antennas & $M_\text{BS} = 50$ \\
\hline
Number of users & $N_\text{UE} = 20$ \\
\hline
Number of RISs & $N_\text{RIS} = 10$ \\
\hline
Number of RIS elements & $M_\text{RIS} = 250$ \\
\hline
Transmit power per subcarrier & $P = 100\,\text{mW}$ \\
\hline
Subcarrier bandwidth & $15\,\text{kHz}$ \\
\hline
AWGN noise power spectral density & $-174\,\text{dBm/Hz}$ \\
\hline
Noise figure & $6\,\text{dB}$ \\
\hline
Path-loss exponent under LOS (NLOS) & $2 \ (4.25)$ \\
\hline
$K$-factor under LOS (NLOS) & $100 \ (3)$ \\
\hline
Distance-dependent LOS-probability & $p_\text{LOS}(d) = e^{-d/50}$ \\
\hline
Shadow fading variance & $10\,\text{dB}$ \\
\hline
Auction initial price & $p_0 = 0.05$ \\
\hline
Auction price increment & $\Delta p = 0.05$ \\
\hline
Budget & $B^{(b)}_0 = 1$ \\
\hline
\end{tabular}
\label{tab:sim_params}
\end{center}
\end{table}

We evaluated the proposed auction-based allocation in a two-cell scenario focusing on the cell-edge region, where users experience the most ambiguous association conditions. We adopt a clustered geometry in which both RISs and users are concentrated close to the cell boundary. This reflects the practical observation that RIS-assisted links provide the strongest benefit when the RIS is positioned near either the transmitter or the receiver, rather than midway between them \cite{Toumi2021}. Specifically, we place $N_\text{BS} = 2$ base stations at opposite ends of a $100\,\text{m}^2$ region of interest and generate correlated clusters of cell-edge users and RISs around the boundary line, as illustrated in Fig.~\ref{fig:plot_positions_RL} for an example RIS allocation obtained with an RL agent at $\beta=3$. The parameters of the simulations are listed in Table~\ref{tab:sim_params}. 

The RL agent undergoes training through episodes. Training proceeds for up to $3 \cdot 10^6$ auction steps, with early stopping applied if no improvement in performance is observed over several evaluation intervals. During the evaluation, results are averaged over 200 macroscopic realizations (positions, path gains), each comprising 20 independent microscopic fading realizations (Rayleigh fading component).

\begin{figure}[t]
\centerline{\includegraphics[width=\columnwidth]{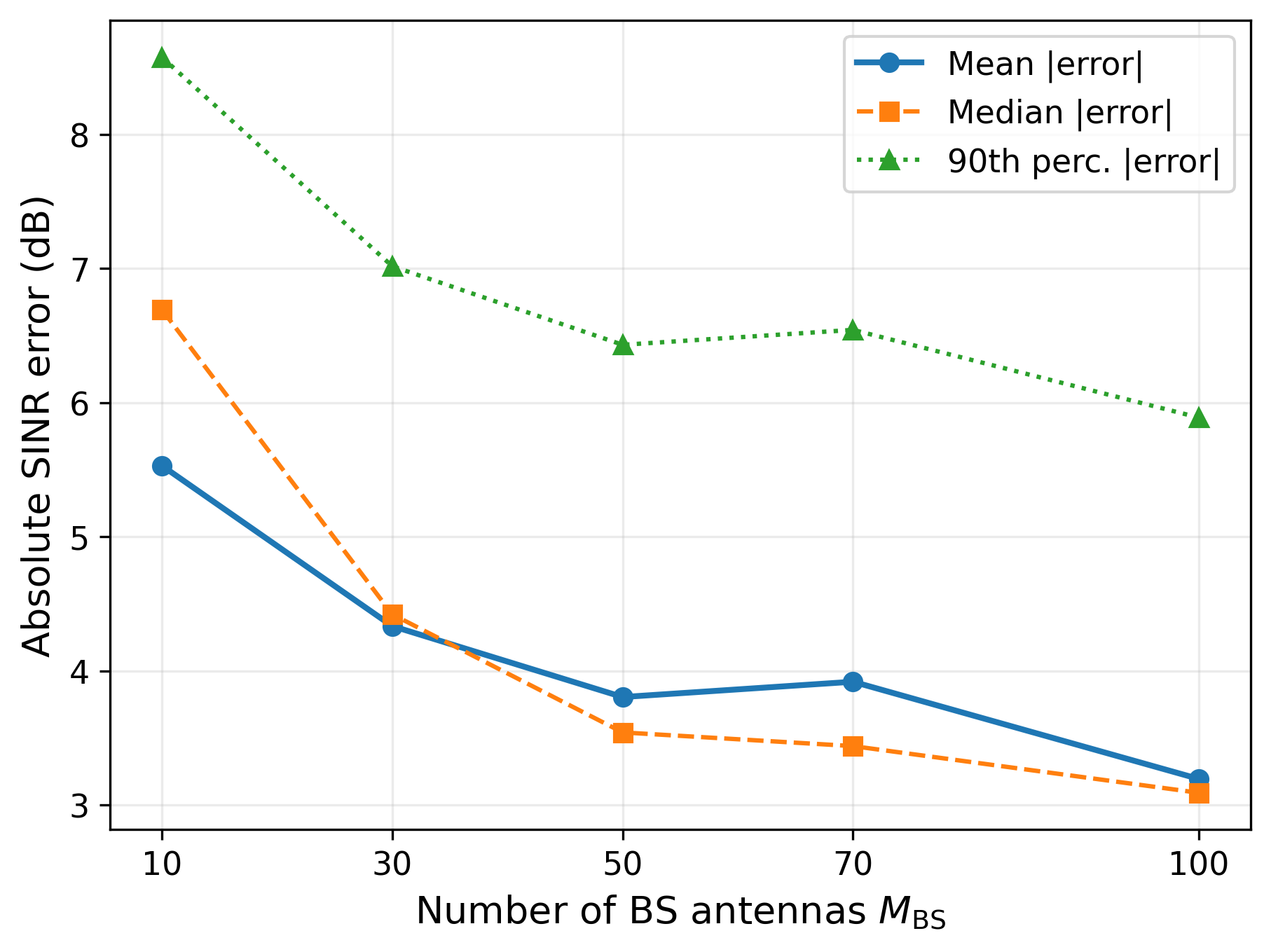}}
\caption{Accuracy of the macroscopic SINR estimation as a function of the number of BS antennas $M_{\mathrm{BS}}$. The figure reports the mean, median, and 90th percentile of the absolute error between the estimated and true SINR values.}
\label{fig:sinr_accuracy}
\end{figure}

To assess the accuracy of the proposed macroscopic SINR estimation, Fig.~\ref{fig:sinr_accuracy} reports multiple statistics of the absolute error between the estimated and true SINR values for different numbers of BS antennas $M_{\mathrm{BS}}$ for a single macroscopic realization. The results reveal a clear trend of decreasing estimation error as $M_{\mathrm{BS}}$ increases, indicating that the macroscopic approximation becomes more reliable for larger arrays. For smaller array sizes, a noticeable estimation mismatch persists.

\begin{figure}[t]
\centerline{\includegraphics[width=\columnwidth]{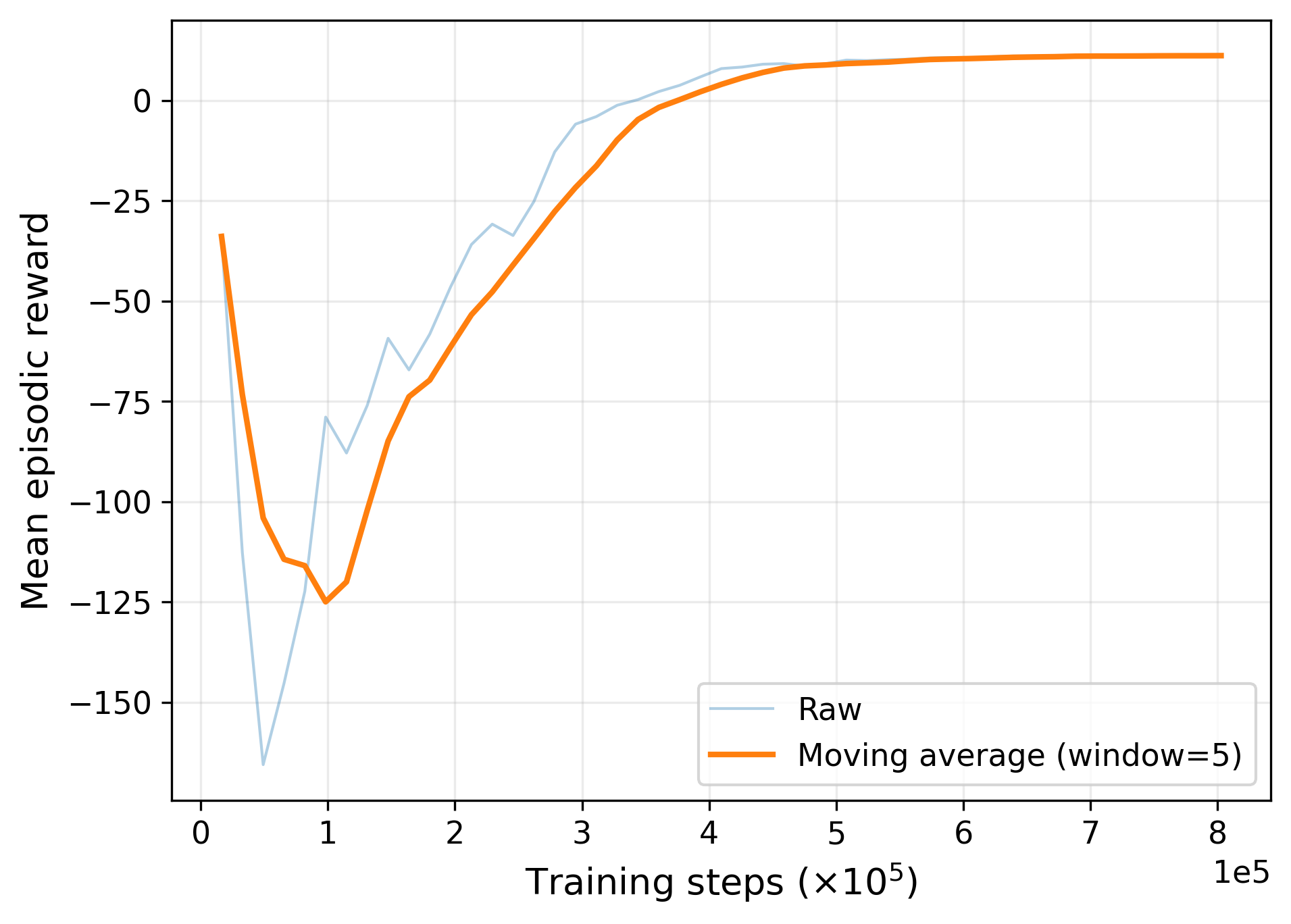}}
\caption{Training convergence of the PPO-based bidding agent with $\beta=2$. The figure shows the evolution of the training reward over environment interaction steps, with a moving-average smoothed curve (window size 5).}
\label{fig:convergence}
\end{figure}

To improve the credibility and reproducibility of the simulation results, we analyze the training dynamics of the DRL-based bidding agent. Fig.~\ref{fig:convergence} shows the evolution of the reward value during learning. Following an initial transient phase with high variability, the reward improves steadily and converges to a stable plateau, indicating that the policy has converged prior to evaluation. A similar convergence behavior is observed for the other considered models, although larger $\beta$ values lead to lower final reward levels.

\begin{figure}[t]
\centerline{\includegraphics[width=\columnwidth]{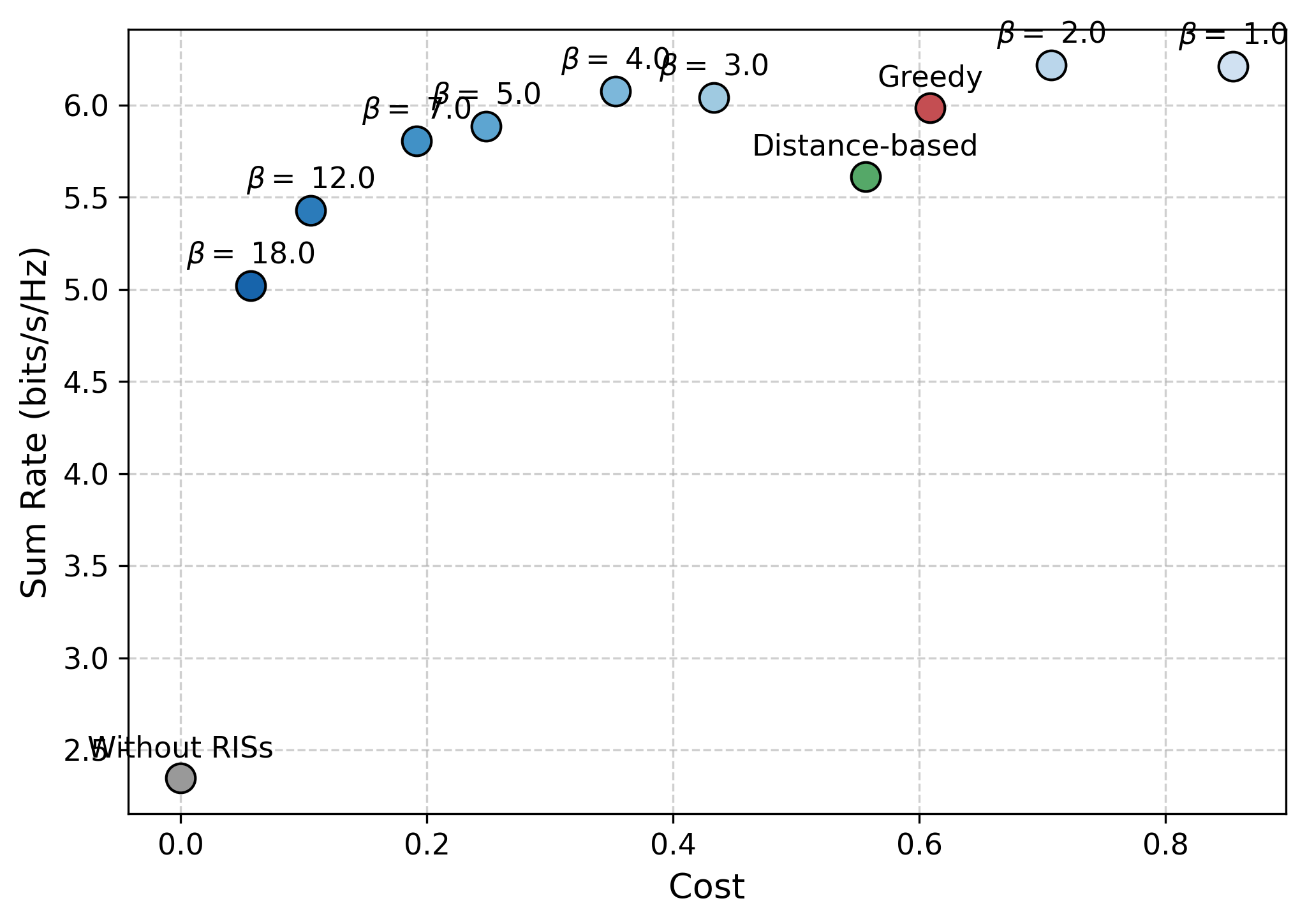}}
\caption{Performance comparison of heuristics and RL-based models using different bid intensity ($\beta$) values. The figure illustrates the trade-off between cost and achievable rate.}
\label{fig:cost_vs_sum_rate}
\end{figure}

Fig.~\ref{fig:cost_vs_sum_rate} illustrates the trade-off between cost and achievable sum rate for the proposed RL-based allocation and the baseline strategies. Three reference points are considered: the two heuristic approaches introduced in Subsection~\ref{IV-A}, and the ``Without RISs'' case in which no RISs are allocated. From this figure, three main observations can be made. First, the absence of RISs leads to significantly lower performance, underlining the importance of RIS-assisted transmission. It should be noted that this effect is overemphasized in our simulations, due to the assumption that there is no direct LOS between users and base stations. Second, compared to the heuristics, several RL models achieve better solutions, i.e., they provide higher sum rates at lower costs. This indicates that the RL agents learn to selectively acquire higher-value RISs, whereas the heuristics tend to bid more aggressively and thus increase the overall cost. Third, the bid intensity parameter $\beta$ effectively regulates the aggressiveness of bidding: larger $\beta$ values correspond to more conservative behavior, resulting in lower costs but also reduced performance. Consequently, $\beta$ provides a practical mechanism to balance the sum rate and the cost.

\begin{figure}[t]
\centerline{\includegraphics[width=\columnwidth]{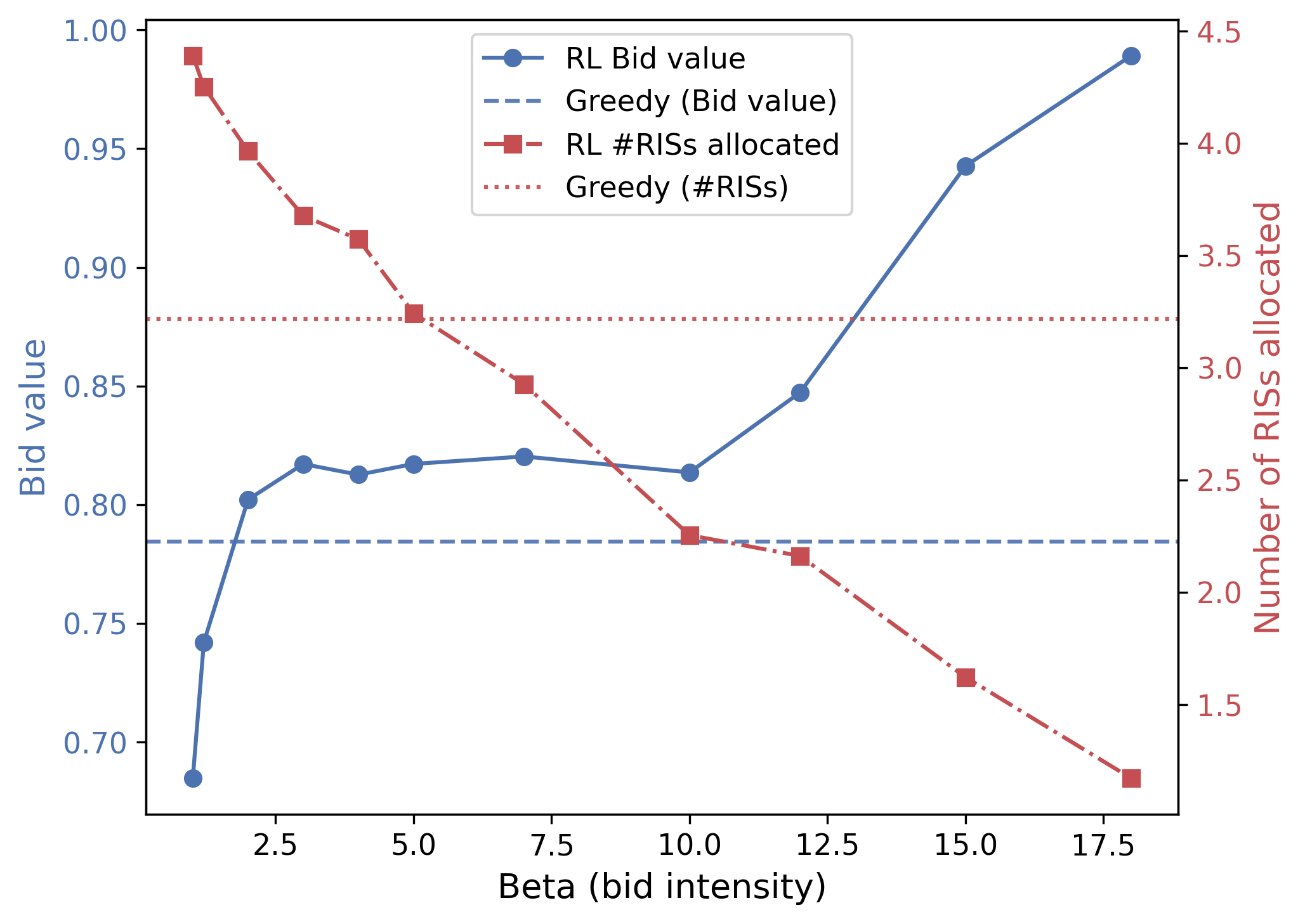}}
\caption{Impact of the bid intensity ($\beta$) parameter on RL bidding behavior. Larger $\beta$ values make agents more selective, resulting in higher average bid values but fewer allocated RISs, while smaller $\beta$ values lead to broader bidding on lower-value RISs.}
\label{fig:beta_effects_with_baselines}
\end{figure}

Fig.~\ref{fig:beta_effects_with_baselines} shows the impact of the bid intensity parameter $\beta$ on the bidding behavior of the RL agents. The left vertical axis indicates the normalized bid values, as defined in~\eqref{eq:value}, with a maximum value of one. As $\beta$ increases, the average bid value also increases, meaning that the agents become more selective and tend to focus on high-value RISs. For small values of $\beta$, the agents are willing to also acquire lower-value RISs, whereas for medium and large values of $\beta$, the bidding is concentrated on RISs of higher utility. The right vertical axis depicts the number of RISs allocated. Here, the opposite trend can be observed: larger $\beta$ values lead to fewer allocated RISs, since the agents adopt a more conservative strategy and place fewer bids overall.

\section{Conclusion}

In this work, we investigated auction-based allocation of reconfigurable intelligent surfaces in a two-cell scenario with clustered deployments at the cell edge. Our results demonstrate that RISs substantially enhance network performance compared to scenarios without RIS assistance. Furthermore, reinforcement learning based bidding strategies outperform heuristic approaches by achieving superior trade-offs, i.e., higher sum rates at lower costs. Finally, we showed that the bid intensity parameter $\beta$ provides an effective control mechanism for adjusting the aggressiveness of bidding, thereby enabling a flexible balance between spectral efficiency and expenditure. These findings highlight the potential of reinforcement learning for efficient and adaptive resource allocation in RIS-assisted networks.

\end{document}